\documentclass[journal,twocolumn]{IEEEtran}
\usepackage{amssymb}
\usepackage{amsfonts}
\usepackage{amsmath}
\usepackage{algorithm}
\usepackage{algorithmic}
\usepackage{subeqnarray}
\usepackage{cases}
\usepackage{color}
\usepackage{bm}
\usepackage{mathrsfs}
\usepackage{graphicx,subfigure,amsmath,amssymb,cite}
\usepackage{float}
\usepackage[switch]{lineno}

\ifCLASSINFOpdf
\else
\fi
\begin{document}

\title{Pilot Optimization and Channel Estimation for Two-way Relaying Network Aided by IRS with Finite Discrete Phase Shifters}

\author{Zhongwen Sun, Xuehui Wang, Siling Feng, Xinrong Guan, Feng Shu, and Jiangzhou Wang, ~\emph{Fellow}~\emph{IEEE}
%\thanks{Copyright (c) 2015 IEEE. Personal use of this material is permitted. However, permission to use this material for any other purposes must be obtained from the IEEE by sending a request to pubs-permissions@ieee.org.}
\thanks{This work was supported in part by the National Natural Science Foundation of China (Nos. 62071234 and 61771244), and the Scientific Research Fund Project of Hainan University under Grant KYQD(ZR)-21008 and KYQD(ZR)-21007.(Corresponding authors: Feng Shu)}
%\thanks{Tong Shen,~Jin Wang,~and Feng Shu are with the School of Electronic and Optical Engineering, Nanjing University of Science and Technology, 210094, CHINA. (Email: shufeng0101 @163.com). }
\thanks{Zhongwen Sun,~Xuehui Wang,~ Siling Feng,~and Feng Shu are with the School of Information and Communication Engineering, Hainan University, Haikou, 570228, China.(Email: shufeng0101@163.com)}
\thanks{Xinrong Guan is with the Communications Engineering College, Army Engineering University of PLA, Nanjing, 210007, China (Email: guanxr@aliyun.com).}
%\thanks{Shuo Zhang is with the National Key Laboratory of Science and Technology on Aerospace Intelligence Control, Beijing Aerospace Automatic Control Institute, Beijing, 100854, China. (Email: gcshuo@163.com)}
\thanks{Jiangzhou Wang is with the School of Engineering, University of Kent, Canterbury CT2 7NT, U.K. (Email: j.z.wang@kent.ac.uk).}
}
\maketitle
\begin{abstract}
In this paper, we investigate the problem of pilot optimization and channel estimation of  two-way relaying network (TWRN) aided by an intelligent reflecting surface (IRS) with finite discrete phase shifters. In a TWRN, there exists a challenging problem that the two cascading channels from User1-to-IRS-to-relay and User2-to-IRS-to-relay and two direct channels from User1-to-relay and User2-to-relay interfere with each other. Via smartly designing the initial phase shifts of IRS and pilot pattern, the two cascading channels are separated over only four pilot sequences  by using simple arithmetic operations like addition and subtraction. Then, the least-squares estimator is adopted to estimate the two cascading channels and two direct channels. The corresponding sum mean square errors (MSE) of channel estimators are derived. By minimizing Sum-MSE, the optimal phase shift matrix of IRS is proved. Then, two special matrices Hadamard and discrete Fourier transform (DFT) matrix is shown to be two optimal training matrices for IRS. Furthermore, the IRS with discrete finite phase shifters is taken into account. Using theoretical derivation and numerical simulations, we find that 3-4 bits phase shifters are sufficient for IRS to achieve a negligible MSE performance loss. More importantly, the Hadamard matrix requires only one-bit phase shifters to achieve the optimal Sum-MSE performance while the DFT matrix requires at least three or four bits to achieve the same performance. Thus, the Hadamard matrix is a perfect choice for channel estimation  using low-resolution phase-shifting IRS.
%We choose the Hadamard matrix as the training pattern instead of Discrete Fourier Transform(DFT) matrix for low-resolution phase shifters of IRS.Also, we perform simulation on the loss caused by quantization of DFT matrix.And we calculate the approximation of the quantization loss.And 3-4 bits are the appropriate numbers of quantization bits.
\end{abstract}
\section{Introduction}
%Recently
%Mobile communication technologies have been developed rapidly in recent years [A-B]. One novel technique
Recently, intelligent reflecting surface (IRS), consisting of many passive reflecting units, attracts a heavy research activities from academia and industry due to its low-cost and low-power consumption. Compared with relay \cite{JPSR}, IRS owns its unique advantages such as no radio frequency chains, real-time reflecting relay and high energy efficiency. IRS has the potential to be applied in fifth-generation (B5G), sixth-generation (6G), and internet of things (IoT)\cite{TWRIS,9205230}. IRS has been investigated for many scenarios in wireless communications, such as physical layer security, directional modulation, beamforming, energy transmission, and covert communications \cite{IRSSTwoEC,IRSdm,ChengIRS,ShiPower,ZhouIRS}. The combination of relay and IRS strikes a good balance among reducing circuit cost, lowering power consumption, and improving spectral efficiency\cite{wang2021beamforming,HRIRS}.
%Recently, in academic research\cite{IRSSTwoEC,IRSdm,ChengIRS,ShiPower},the beamforming of transmitter ,the phase shift of IRS,and directional modulation in a IRS-aided system has been extensively studied.

%Accurate channel estimation is required for the above scenarios.
Accurate channel estimation is critical for mobile communication systems \cite{ZHOUYQOFCDM}. There has been some research work on channel estimation for IRS-aided wireless communication \cite{jensenDFT,zhengCE2,guananchor,HuTTCE}. In \cite{jensenDFT}, a DFT matrix was selected as the training phase shift matrix for the minimum variance unbiased (MVU) estimator. In \cite{zhengCE2}, the selection of DFT matrix was extended to the RIS-aided single input single output (SISO) orthogonal frequency division multiplexing access (OFDMA) multi-user scenario with innovative pilot pattern to accommodate more users than conventional pilot pattern. In \cite{guananchor}, an anchor-assisted two-phase channel estimation scheme was proposed, where two anchors was placed near the IRS for reducing the overhead of multi-user channel estimation. In \cite{HuTTCE}, a two-timescale channel estimation structure was proposed for reducing the pilot overhead with a dual-link pilot transmission.

% The DFT matrix estimation was extended to the RIS-aided orthogonal frequency division multiplexing(OFDM) signal-user in \cite{zhengCE}. In \cite{zhengCE2}, the above scenario was extended to a multi-user scenario with innovative pilot pattern to accommodate more users than conventional pilot pattern.
%Current research focuses on the use of DFT matrix as training matrices,which often considers that each element of IRS is infinite phase shifter.
However, infinite phase shifter or high-resolution phase shifter can lead to higher cost on hardware. There has been some literature concerning IRS equipped with low-resolution phase shifters. In \cite{YOUDIS1}, a least squares (LS) channel estimator for an IRS-aided single user SISO system was proposed and a low-complexity passive beamforming algorithm was designed based on the channel estimation. In this paper, we make an insight investigation of the problem of pilot optimization and channel estimation of  two-way relaying network (TWRN) aided by IRS with finite-phase shifters. The main contributions of this paper are summarized as follows:
\begin{itemize}
  \item To improve the performance and reduce the computational complexity,  a perfect pilot pattern is proposed for an IRS-aided TWRN. Using such a pattern,  four coupled channels including two cascading channels and two direct channels are separated %smartly and
      completely via some simple arithmetic operation like add and subtract. Then, via LS rule, the four channels may be independently estimated.  Finally, the optimal training matrix is derived by minimizing sum mean square errors (MSE), and proves the fact that the training matrix is  a unitary matrix times a constant. With constant-modulus constraint, the Hadamard and DFT matrices are shown to be the optimal choice for the phase matrix of IRS with infinite-phase shifters.
  \item For an IRS with finite-phase shifters,  the quantization performance loss  factor is defined and derived with  DFT matrix as an example.  In general, a DFT matrix requires $2\log_2N$ bits to achieve a channel estimator  without  performance loss, where $N$ denotes the number of points of DFT being  taken to be the number $M$ of IRS elements for the convenience of deriving below. According to the performance loss factor, $3\sim4$ bits are sufficient for a DFT matrix to realize an omitted Sum-MSE performance loss. In particular, a Hadamard matrix requires only one-bit. This makes  Hadamard matrix more attractive than DFT one, especially, in the scenario of IRS employing low-cost  and low-resolution  phase shifters.
      %Simulation results show that MSE decreases as the number of phase shifter bit rises and eventually approximates the scene without quantization. The downward trend of MSE is constant regardless of the signal-to-noise ratio. A theoretical derivation is performed for evaluate the loss during the quantization. It is concluded that 3-4 bits are the appropriate phase shifters based on the results of the derivation and simulation. Simulation results show that Hadamard matrix and DFT matrix perform consistently and much better than random phase shift matrix. However, the choice of DFT matrix requires the high-resolution phase shifters when the number of IRS elements is high, while Hadamard matrix requires only one-bit phase shifters.
\end{itemize}
The remainder is organized as follows. Section \ref{model} describes the system model. Channel estimation, pilot design, and performance analysis of quantization error are presented in Section \ref{method}. In Section \ref{simulation}, numerical simulations are conducted, and we conclude in Section \ref{conclusion}.

\noindent \emph{Notations}: Scalars, vectors and matrices are respectively represented by letters of lower case, bold lower case, and bold upper case. $(\cdot)^*$, $(\cdot)^H$, $(\cdot)^T$ stand for matrix conjugate, conjugate transpose, and transpose, respectively. $\mathbb{E}\{\cdot\}$, $\|\cdot\|_F$and $\text{tr}\{\cdot\}$ denote expectation operation, Frobenius norm, the trace of a matrix, respectively. $\text{vec}(\cdot)$ denotes vector operator. $\odot$ denotes Hadamard product.
\section{System Model}\label{model}
\begin{figure}[htb]
\centering
\includegraphics[width=0.28\textwidth]{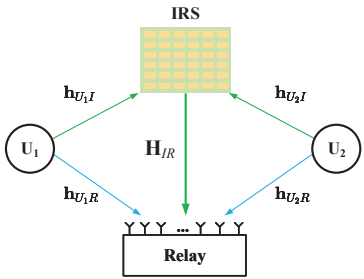}
\centering
\caption{Diagram block for TWRN assisted by IRS.}
\label{fig1}
\end{figure}

Fig.~\ref{fig1} sketches a TWRN system with two users. It is assumed  User1 and User2 is blocked and there is no direct link between them. But they can transmit signals to each other with a half-duplex relay and IRS.  User1, User2, relay station, IRS are denoted by $\text{U}_1$, $\text{U}_2$, R and I, respectively. Relay  and IRS are employed with $K$ antennas and  $M$ reflecting elements. Without loss of generality, it is assumed that all channels follow Rayleigh fading. Channel frequency responses (CFR) of $\text{U}_1$ $\rightarrow$ I, I $\rightarrow$ R, $\text{U}_1$ $\rightarrow$ R, $\text{U}_2$ $\rightarrow$ I, and $\text{U}_2$ $\rightarrow$ R are denoted by $ \textbf{h}_{{U}_{1}I}\in\mathbb{C}^{M \times 1}$, $\textbf{H}_{IR}\in\mathbb{C}^{K \times M}$, $\textbf{h}_{{U}_{1}R}\in\mathbb{C}^{K \times 1}$, $\textbf{h}_{{U}_{2}I}\in\mathbb{C}^{M \times 1}$, and $\textbf{h}_{{U}_{2}R}\in\mathbb{C}^{K \times 1}$, respectively.

User1 and User2 transmit their symbols to the relay simultaneously, and the receive signal at relay can be written as
\begin{align}\label{y1}
\mathbf{y}&=\sqrt{{P}_{{U}_{1}}}(\mathbf{H}_{IR} \bm{\Theta}\mathbf{h}_{{U}_{1}I}+\mathbf{h}_{{U}_{1}R})x_{{U}_{1}}  \nonumber \\
&\quad + \sqrt{P_{{U}_{2}}}(\mathbf{H}_{IR} \bm{\Theta}\mathbf{h}_{{U}_{2}I} +\mathbf{h}_{{U}_{2}R})x_{{U}_{2}}+\mathbf{w}
\end{align}
where $\mathbf{y} \in \mathbb{C}^{K \times 1} $ denotes the received signal vector, $x_{{U}_{1}}$ and $x_{{U}_{2}}$ denote the transmitted symbol from user1 and user2, respectively, $\mathbf{w} \in \mathbb{C}^{K \times 1}$ denotes the receive additive white Gaussian noise with $\mathbf{w}\thicksim \mathcal{CN} (0,{\sigma}^2 I)$, and the diagonal matrix $\bm{\Theta}$ is the phase matrix of IRS defined as
\begin{align}\label{irs}
\bm{\Theta}=\text{diag}({\alpha}_m e^{-j{\vartheta}_m})=\text{diag}({\theta}_m), m=1,2,\ldots,M
\end{align}
where ${\alpha}_m \in(0,1]$ and ${\vartheta}_m \in [0,2\pi]$ ,$m=1,\ldots,M$ stand for the amplitude value and phase shift value of the $m$-th reflection elements, respectively. For simplicity, all of the amplitude values ${\alpha}_m$ are set to 1.  In (\ref{y1}), the channel product $\bm{\Theta}\mathbf{h}_{SI}$ can be rewritten as follows
\begin{align}\label{mT}
\bm{\Theta}\mathbf{h}_{{U}_{1}I}
&=[{\theta}_1 h_{{U}_{1}I}(1),{\theta}_2 h_{{U}_{1}I}(2),\ldots,{\theta}_M h_{{U}_{1}I}(M)]^T \nonumber \\
&=\mathbf{H}_{{U}_{1}I}\mathbf{\bm{\theta}}
\end{align}
where $\mathbf{H}_{{U}_{1}I}=\text{diag}(\mathbf{h}_{{U}_{1}I})$ and
$\bm{\theta}\triangleq [{\theta}_1,{\theta}_2,\ldots,{\theta}_M]^T=[{\alpha}_1 e^{-j{\vartheta}_1},{\alpha}_2 e^{-j{\vartheta}_2},\ldots,{\alpha}_M e^{-j{\vartheta}_M}]^T$. Similarly, we can get $\bm{\Theta}\mathbf{h}_{{U}_{2}I} = \mathbf{H}_{{U}_{2}I}\mathbf{\bm{\theta}}$ where $\mathbf{H}_{{U}_{2}I}=\text{diag}(\mathbf{h}_{{U}_{2}I})$.
By denoting the cascading channels as $\mathbf{H}_{{U}_{1}IR}=\mathbf{H}_{IR}\mathbf{H}_{{U}_{1}I}\in \mathbb{C}^{K \times M}$ and $\mathbf{H}_{{U}_{2}IR}=\mathbf{H}_{IR}\mathbf{H}_{{U}_{2}I}\in \mathbb{C}^{K \times M}$, the received signal (\ref{y1}) can be rewritten as
\begin{align}\label{y2}
\mathbf{y}&=\sqrt{P_{{U}_{1}}}(\mathbf{H}_{{U}_{1}IR}\bm{\theta}+\mathbf{h}_{{U}_{1}R})x_{{U}_{1}} \nonumber \\
&\quad + \sqrt{P_{{U}_{2}}}(\mathbf{H}_{{{U}_{2}}IR}\bm{\theta}+\mathbf{h}_{{{U}_{2}}R})x_{{U}_{2}}+\mathbf{w}.
\end{align}
\section{Proposed IRS channel estimator, pilot pattern and performance loss analysis}\label{method}
In this section,  the LS channel estimator is proposed. Then,  the optimal pilot pattern and phase shift training matrix are derived. Furthermore, in the scenario with a finite-phase-shifter IRS, the MSE performance loss factor is derived  and analyzed due to the effect of quantization error.
\subsection{Proposed LS channel estimator}
\begin{figure}[htb]
\centering
\includegraphics[width=0.40\textwidth]{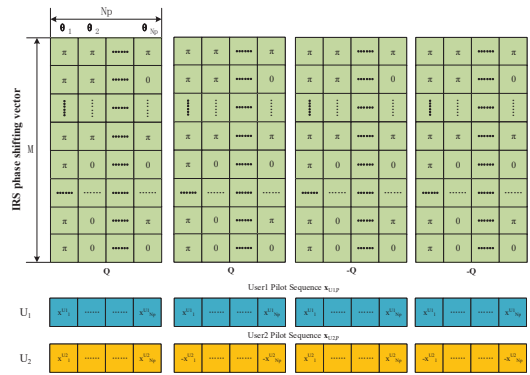}
%\captionsetup{justification=centering}
\caption{Proposed pilot pattern.}
\label{fig2}
\end{figure}

Fig.~\ref{fig2} shows the proposed pilot pattern. Here, user1 and user2 firstly send their pilot sequences  $\mathbf{x}_{{U}_{1}}=[x_{{U}_{1}}(1),...,x_{{U}_{1}}(N_P)]^T$  and $\mathbf{x}_{{U}_{2}}=[x_{{U}_{2}}(1),...,x_{{U}_{2}}(N_P)]^T$  of   $N_P$  symbols. Each symbol in the pilot sequence is related to a different IRS phase configuration $\bm{\theta}_i$, where $i= 1,2,\ldots,N_P$. For user1 and user2, they transmit four continuous pilot sequences. Four sequences of the former are identical while the latter changes those sequence signs alternatively. Phase shift matrix $\mathbf{Q}$ change its signs in the latter two sequences.

In the first pilot sequence period, the received signal can be expressed as 
\begin{align}\label{hse1}
&\mathbf{Y}_1=[\mathbf{y}_1,\mathbf{y}_2,\ldots,\mathbf{y}_{N_P}] = \sqrt{P_{{U}_{1}}}\mathbf{H}_{{U}_{1}IR}\mathbf{Q}\mathbf{X}_{{U}_{1}} + \sqrt{P_{{U}_{1}}} \nonumber \\
%&=\sqrt{P_S}\mathbf{H}_{SIR}[\bm{\theta}_1 x_{S,P}(1),\bm{\theta}_2 x_{S,P}(2),\ldots,\bm{\theta}_{N_P} x_{S,P}(N_P)] \nonumber \\
%& + \sqrt{P_S}\mathbf{h}_{SR}[x_{S,P}(1), x_{S,P}(2),\ldots,x_{S,P}(N_P)] \nonumber \\
%& + \sqrt{P_D}\mathbf{H}_{DIR}[\bm{\theta}_1 x_{D,P}(1),\bm{\theta}_2 x_{D,P}(2),\ldots,\bm{\theta}_{N_P} x_{D,P}(N_P)] \nonumber \\
%& + \sqrt{P_D}\mathbf{h}_{DR}[x_{S,D}(1), x_{S,D}(2),\ldots,x_{S,D}(N_P)]+ \mathbf{W}_1  \nonumber \\
&\mathbf{h}_{{U}_{1}R} \mathbf{x}^T_{{U}_{1}}+ \sqrt{P_{{U}_{2}}}\mathbf{H}_{{U}_{2}IR}\mathbf{Q}\mathbf{X}_{{U}_{2}} + \sqrt{P_{{U}_{2}}}\mathbf{h}_{{U}_{2}R}\mathbf{x}^T_{{U}_{2}} + \mathbf{W}_1
\end{align}
where $\mathbf{Q}=[\bm{\theta}_1,\bm{\theta}_2,\ldots,\bm{\theta}_{Np}]$, $\mathbf{X}_{{U}_{1}}=\text{diag}(\mathbf{x}_{{U}_{1}})$, and $\mathbf{X}_{{U}_{2}}=\text{diag}(\mathbf{x}_{{U}_{2}})$.
Similarly, we have three receive signal matrices corresponding to the last three pilot sequences as follows 
%In the second pilot sequence period, let $\mathbf{x}^{(2)}_{D,P}=-\mathbf{x}_{D,P}$. We can separate the channel of S and D as follows,
\begin{align}\label{hse2}
\mathbf{Y}_2&=\sqrt{P_{{U}_{1}}}\mathbf{H}_{{U}_{1}IR}\mathbf{Q}\mathbf{X}_{{U}_{1}} + \sqrt{P_{{U}_{1}}}\mathbf{h}_{{U}_{1}R}\mathbf{x}^T_{{U}_{1}}  \nonumber \\
&\quad - \sqrt{P_{{U}_{2}}}\mathbf{H}_{{U}_{2}IR}\mathbf{Q}\mathbf{X}_{{U}_{2}} - \sqrt{P_{{U}_{2}}}\mathbf{h}_{{U}_{2}R}\mathbf{x}^T_{{{U}_{1}}} + \mathbf{W}_2,
\end{align}
%At the third pilot training time slot, let $\vartheta^(3)_m = \vartheta_m +\pi$, which makes $\bm{\theta}^{(3)}_i = - \bm{\theta}_i$ .
\begin{align}\label{hse3}
\mathbf{Y}_3&= -\sqrt{P_{{U}_{1}}}\mathbf{H}_{{U}_{1}IR}\mathbf{Q}\mathbf{X}_{{U}_{1}} + \sqrt{P_{{U}_{1}}}\mathbf{h}_{{U}_{1}R}\mathbf{x}^T_{{U}_{1}}  \nonumber \\
&\quad - \sqrt{P_{{U}_{2}}}\mathbf{H}_{{U}_{2}IR}\mathbf{Q}\mathbf{X}_{{U}_{2}} + \sqrt{P_{{U}_{2}}}\mathbf{h}_{{U}_{2}R}\mathbf{x}^T_{{{U}_{1}}} + \mathbf{W}_3,
\end{align}
and
%At the fourth pilot training time slot , the IRS configuration is the same as the third time slot, and $\mathbf{x}^{(4)}_{D,P}=-\mathbf{x}_{D,P}$.
\begin{align}\label{hse4}
\mathbf{Y}_4&= -\sqrt{P_{{U}_{1}}}\mathbf{H}_{{U}_{1}IR}\mathbf{Q}\mathbf{X}_{{U}_{1}} + \sqrt{P_S}\mathbf{h}_{SR}\mathbf{x}^T_{S,P}  \nonumber \\
&\quad + \sqrt{P_{{U}_{2}}}\mathbf{H}_{{U}_{2}IR}\mathbf{Q}\mathbf{X}_{{U}_{2}} - \sqrt{P_{{U}_{2}}}\mathbf{h}_{{U}_{2}R}\mathbf{x}^T_{{{U}_{1}}} + \mathbf{W}_4.
\end{align}
Observing the above four equations, due to their symmetric property, we have readily obtained the following four individual cascading and direct channels equations
\small{
\begin{align}\label{hse5}
\tilde{\mathbf{Y}}_1=\mathbf{Y}_1 + \mathbf{Y}_2  &+ \mathbf{Y}_3 + \mathbf{Y}_4= 4\sqrt{P_{{U}_{1}}}\mathbf{h}_{{U}_{1}R}\mathbf{x}^T_{{U}_{1}} \nonumber \\
&\quad + \mathbf{W}_1 + \mathbf{W}_2 + \mathbf{W}_3 + \mathbf{W}_4
\end{align}
\begin{align}\label{hse6}
\tilde{\mathbf{Y}}_2=\mathbf{Y}_1 - \mathbf{Y}_2 &+ \mathbf{Y}_3 - \mathbf{Y}_4 = 4\sqrt{P_{{U}_{2}}}\mathbf{h}_{{U}_{2}R}\mathbf{x}^T_{{U}_{2}} \nonumber \\
&\quad + \mathbf{W}_1 - \mathbf{W}_2 + \mathbf{W}_3 - \mathbf{W}_4
\end{align}
}
\small{
\begin{align}\label{hse7}
\tilde{\mathbf{Y}}_3=\mathbf{Y}_1 + \mathbf{Y}_2  &- \mathbf{Y}_3 - \mathbf{Y}_4= 4\sqrt{P_{{U}_{1}}}\mathbf{H}_{{U}_{1}IR}\mathbf{Q}\mathbf{X}_{{U}_{1}} \nonumber \\
&\quad + \mathbf{W}_1 + \mathbf{W}_2 - \mathbf{W}_3 - \mathbf{W}_4
\end{align}
\begin{align}\label{hse8}
\tilde{\mathbf{Y}}_4=\mathbf{Y}_1 - \mathbf{Y}_2& - \mathbf{Y}_3 + \mathbf{Y}_4 = 4\sqrt{P_{{U}_{2}}}\mathbf{H}_{{U}_{2}IR}\mathbf{Q}\mathbf{X}_{{U}_{2}} \nonumber \\
&\quad + \mathbf{W}_1 - \mathbf{W}_2 - \mathbf{W}_3 + \mathbf{W}_4
\end{align}
}
where  the number of columns of  matrix $\mathbf{Q}$ is chosen to be greater than or equal to the number of its rows in order to ensure that $\mathbf{Q}$ is invertible, i.e. $N_P \geq M$.  To reduce the estimation overheads, it is assumed that $N_P=M$. Multiplying (\ref{hse5})  by  $\mathbf{x}^*_{{U}_{1}}$ from the right gives
\begin{align}\label{ses1}
\tilde{\mathbf{Y}}_1 \mathbf{x}^*_{{U}_{1}P} &= 4\sqrt{P_{{U}_{1}}}\mathbf{h}_{{U}_{1}R}\mathbf{x}^T_{{U}_{1},P}\mathbf{x}^*_{{U}_{1}} \nonumber \\
&\quad + (\mathbf{W}_1 + \mathbf{W}_2 + \mathbf{W}_3 + \mathbf{W}_4)\mathbf{x}^*_{{U}_{1}}
\end{align}
which yields the LS estimation of $\mathbf{h}_{{U}_{1}R}$.
\begin{align}\label{e1}
  \hat{\mathbf{h}}_{{U}_{1}R}= \tfrac{\tilde{\mathbf{Y}}_1 \mathbf{x}^*_{{U}_{1}}}{4\sqrt{P_{{U}_{1}}}\mathbf{x}^T_{{U}_{1}}\mathbf{x}^*_{{U}_{1}}}.
\end{align}
Similarly,
\begin{align}\label{e2}
  \hat{\mathbf{h}}_{{U}_{2}R}= \tfrac{\tilde{\mathbf{Y}}_2 \mathbf{x}^*_{{U}_{2}}}{4\sqrt{P_{{U}_{1}}}\mathbf{x}^T_{{U}_{2}}\mathbf{x}^*_{{U}_{2}}}.
\end{align}
Now, letting us turn to the cascading channels, performing vec operation on two sides of Eq.(\ref{hse7}) forms
\begin{align}\label{ses2}
\text{vec}(\tilde{\mathbf{Y}}_3)&=4\sqrt{P_{{U}_{1}}}(\overbrace{(\mathbf{Q} \mathbf{X}_{{U}_{1}})^T \otimes \mathbf{I}_K}^{\mathbf{A}_{{U}_{1}}})\text{vec}(\mathbf{H}_{{U}_{1}IR}) \nonumber \\
& \quad + \text{vec}(\mathbf{W}_1 + \mathbf{W}_2 - \mathbf{W}_3 - \mathbf{W}_4)
\end{align}
which gives the  LS estimator of $\mathbf{H}_{{U}_{1}IR}$ as follows
\begin{align}\label{e3}
\text{vec}(\hat{\mathbf{H}}_{{U}_{1}IR})=\tfrac{\mathbf{A}_{{U}_{1}}^{-1}\text{vec}(\tilde{\mathbf{Y}}_3)}{4\sqrt{P_{{U}_{1}}}}.
\end{align}
Similarly, we have
\begin{align}\label{e4}
\text{vec}(\hat{\mathbf{H}}_{{U}_{2}IR})=\tfrac{\mathbf{A}_{{U}_{2}}^{-1}\text{vec}(\tilde{\mathbf{Y}}_4)}{4\sqrt{P_{{U}_{2}}}}.
\end{align}
Given  $\mathbb{E}\{\mathbf{x}^T_{{U}_{1}}\mathbf{x}^*_{{U}_{1}}\}=M$, the MSE of estimating $\mathbf{h}_{{U}_{1}R}$ is 
\begin{align}\label{mse1}
\epsilon_1 &= \tfrac{1}{4M} \mathbb{E} \{ \| \hat{\mathbf{h}}_{{U}_{1}R} -  \mathbf{h}_{{U}_{1}R} \|^{2}_F\}  \nonumber \\
&=\tfrac{1}{4M} \mathbb{E} \{ \| \tfrac{\tilde{\mathbf{Y}}_1 \mathbf{x}^*_{{U}_{1}}}{4\sqrt{P_S}\mathbf{x}^T_{{U}_{1}}\mathbf{x}^*_{{U}_{1}}} -  \mathbf{h}_{{U}_{1}R} \|^{2}_F\} \nonumber \\
&= \tfrac{1}{4M} \mathbb{E} \{ \| \tfrac{(4\sqrt{P_{{U}_{1}}}\mathbf{h}_{SR}\mathbf{x}^T_{{U}_{1}}+ \mathbf{W}_1 + \mathbf{W}_2 + \mathbf{W}_3 + \mathbf{W}_4) \mathbf{x}^*_{{U}_{1}}}{ 4\sqrt{P_{{U}_{1}}}\mathbf{x}^T_{{U}_{1}}\mathbf{x}^*_{{U}_{1}}} -  \mathbf{h}_{{U}_{1}R} \|^{2}_F\} \nonumber \\
&= \tfrac{1}{4M} \mathbb{E} \{ \|\tfrac {(\mathbf{W}_1 + \mathbf{W}_2 + \mathbf{W}_3 + \mathbf{W}_4)\mathbf{x}^*_{{U}_{1}}}{4\sqrt{P_{{U}_{1}}}\mathbf{x}^T_{{U}_{1}}\mathbf{x}^*_{{U}_{1}}} \|^{2}_F\} \nonumber \\
&= \tfrac{1}{64 P_{{U}_{1}} M^3} \mathbb{E} \{\|(\mathbf{W}_1 + \mathbf{W}_2 + \mathbf{W}_3 + \mathbf{W}_4)\mathbf{x}^*_{{U}_{1}}\|^{2}_F\} \nonumber \\
&=\tfrac{{\sigma}^2}{16 P_{{U}_{1}} M^2}
\end{align}
Similarly, we have the remaining MSEs 
\begin{align}\label{mse2}
\epsilon_2=\tfrac{{\sigma}^2}{16 P_{{U}_{2}} M^2},
\end{align}
\begin{align}\label{mse3}
&\epsilon_3 = \tfrac{1}{4M}\mathbf{E} \{ \| \text{vec}(\hat{\mathbf{H}}_{{U}_{1}IR}) - \text{vec}(\mathbf{H}_{{U}_{1}IR}) \|^{2}_F \} \nonumber \\
&=\tfrac{1}{4M}\mathbf{E} \{ \|  \tfrac{\mathbf{A}_{{U}_{1}}^{-1}\text{vec}(\tilde{\mathbf{Y}}_3)}{4\sqrt{P_{{U}_{1}}}} - \text{vec}(\mathbf{H}_{{U}_{1}IR}) \|^{2}_F \} \nonumber \\
&=\tfrac{1}{4M}\mathbf{E} \{ \|  \tfrac{\mathbf{A}_{{U}_{1}}^{-1}(4\sqrt{P_{{U}_{1}}}\mathbf{A}_{{U}_{1}} \text{vec}(\mathbf{H}_{{U}_{1}IR})+\text{vec}(\mathbf{W}_1 + \mathbf{W}_2 - \mathbf{W}_3 - \mathbf{W}_4))}{4\sqrt{P_{{U}_{1}}}} \nonumber \\
& - \text{vec}(\mathbf{H}_{{U}_{1}IR}) \|^{2}_F \} =\tfrac{1}{4M}\mathbf{E} \{ \|  \tfrac{\mathbf{A}_{{U}_{1}}^{-1}\text{vec}(\mathbf{W}_1 + \mathbf{W}_2 - \mathbf{W}_3 - \mathbf{W}_4)}{4\sqrt{P_{{U}_{1}}}}\|^{2}_F \} \nonumber \\
&=\tfrac{4{\sigma}^2 \text{tr}((\mathbf{A}_{{U}_{1}}^{-1})^H \mathbf{A}_{{U}_{1}}^{-1})}{64 P_{{U}_{1}} M}  = \tfrac{{\sigma}^2 \text{tr}((\mathbf{A}_{{U}_{1}}^{-1})^H \mathbf{A}_{{U}_{1}}^{-1})}{16 P_{{U}_{1}} M},
\end{align}
and
\begin{align}\label{mse4}
\epsilon_4= \tfrac{{\sigma}^2 \text{tr}((\mathbf{A}_{{U}_{2}}^{-1})^H \mathbf{A}_{{U}_{2}}^{-1})}{16 P_{{U}_{2}} M},
\end{align}
which directly yields the Sum-MSE as follows
\begin{align}\label{mse5}
\epsilon =\epsilon_1 +\epsilon_2 +\epsilon_3 +\epsilon_4.
\end{align}
\subsection{Pilot optimization of minimizing Sum-MSE}
%\begin{align}\label{mse6}
%\epsilon_3 =\tfrac{{\sigma}^2 \text{tr}((\mathbf{A}_S^{-1})^H \mathbf{A}_S^{-1})}{16 P_S M}
%\end{align}
Since from Eq. (\ref{mse3}),
\begin{align}\label{invAs}
\mathbf{A}_{{U}_{1}}^{-1}=\overbrace{(\mathbf{Q}^T)^{-1}}^{\mathbf{B}} \mathbf{X}^{-1}_{{U}_{1}} \otimes \mathbf{I}_K,
\end{align}
we have
\begin{align}\label{tr1}
&\text{tr}\{(\mathbf{A}_{{U}_{1}}^{-1})^H \mathbf{A}_{{U}_{1}}^{-1}\}=\text{tr}\{ ((\mathbf{X}^{-1}_{{U}_{1}})^H\mathbf{B}^{H}) \otimes \mathbf{I}_K) (\mathbf{B} \mathbf{X}^{-1}_{{U}_{1}}\otimes \mathbf{I}_K) \} \nonumber \\
&=\text{tr}\{(\mathbf{X}^{-1}_{{U}_{1}})^H \mathbf{B}^{H} \mathbf{B}\mathbf{X}^{-1}_{{U}_{1}} \otimes \mathbf{I}_K)  \} \nonumber \\
&=K\text{tr}\{(\mathbf{X}^{-1}_{{U}_{1}})^H\mathbf{B}^{H}\mathbf{B} \mathbf{X}^{-1}_{{U}_{1}}\} =K\text{tr}\{(\mathbf{X}_{{U}_{1}}^H\mathbf{X}_{{U}_{1}})^{-1}\mathbf{B}^{H}\mathbf{B} \} \nonumber \\
&=K\text{tr}\{(\mathbf{X}_{{U}_{1}}^H\mathbf{X}_{{U}_{1}})^{-1}(\mathbf{Q}^{T}\mathbf{Q}^{*})^{-1} \}
\end{align}
Minimizing Eq. (\ref{tr1}) can be decomposed into the following two sub-problems
\begin{align}\label{problem1}
\text{(P1.1)}:\quad & \underset{\mathbf{Q}}{\min} \ \text{tr}\{(\mathbf{Q}^{T}\mathbf{Q}^{*})^{-1}\} \nonumber \\
& \text{ s.t. } |\theta_t(m)|=1, t=1,2,\ldots,M ,\nonumber \\
& \quad \quad \quad \quad \quad \quad \quad m =1,2,\ldots,M,
\end{align}
and
\begin{align}\label{problem2}
\text{(P1.2)}:\quad & \min K(\text{tr}( (\mathbf{X}^{-1}_{{U}_{1}})^H \mathbf{X}^{-1}_{{U}_{1}})) \nonumber \\
& \text{ s.t. } \tfrac{\text{tr}\{\mathbf{X}_{{U}_{1}}\mathbf{X}^H_{{U}_{1}}\} }{N_P} = 1
\end{align}
%\begin{align}\label{problem1}
%& \min K(\text{tr}((\mathbf{X}^{-1}_{S,P})^H\mathbf{X}^{-1}_{S,P}) + \text{tr}((\mathbf{X}^{-1}_{D,P})^H\mathbf{X}^{-1}_{D,P})) \nonumber \\
%& \text{ s.t. } \tfrac{\text{tr}(\mathbf{X}_{S,P}\mathbf{X}^H_{S,P}) +\text{tr}((\mathbf{X}^{-1}_{D,P})^H\mathbf{X}^{-1}_{D,P})}{2N_P} = 1
%\end{align}

%For problem (P1.1),we perform a singular value decomposition of $\mathbf{Q}^{*}\mathbf{Q}^{T}$ . We can obtain that
%\begin{align}\label{SVD}
%(\mathbf{Q}^{*}\mathbf{Q}^{T})^{-1}=(\mathbf{U}\bm{\Lambda} \mathbf{U}^H)^{-1} = \mathbf{U} \bm{\Lambda}^{-1}\mathbf{U}^{H}
%\end{align}
%
%The problem (P1.1) can be transformed into
%\begin{align}\label{P1.1T}
%& \underset{\bm{\Lambda}}{\min} \  \text{tr}\{\mathbf{U} \bm{\Lambda}^{-1}\mathbf{U}^{H}\} \nonumber \\
%& \text{ s.t. } \text{tr}\{\bm{\Lambda}\} = 1
%\end{align}
%
%Note that $\bm{\Lambda}$ is a diagonal matrix , we can obtain that
%\begin{align}\label{P1.1T2}
%\text{ s.t. } \text{tr}\{\bm{\Lambda}\} = \sum_{i=1}^{M} \lambda_i
%\end{align}
%where $\lambda_i = \bm{\Lambda}_{ii}$,$i=1,2,\ldots,M$. Obviously,the minimum can be obtained when all $\lambda_i$ are equal.
%
%the optimal reflection matrix to minimize the trace should satisfy $\mathbf{Q}^H\mathbf{Q}=M\mathbf{I}_{M}$,which makes $\mathbf{Q}$ a orthogonal matrix.The DFT matrix can satisfy the constant mode constraint.

According to \cite{jensenDFT,zhengCE2,ShuTVTCE}, we have $\|X_{{U}_{1}}(i) \|^2=\|X_{{U}_{2}}(i) \|^2=1, \forall i \in \{1,2,\ldots,N_P\}$ and $\mathbf{Q}^H\mathbf{Q}=M\mathbf{I}_{M}$.
Finally, the minimum of Sum-MSE $\epsilon$ can be given by 
\begin{align}\label{minimse}
\epsilon_{min} =\tfrac{K\sigma^2}{16P_{{U}_{1}} M} + \tfrac{K\sigma^2}{16P_{{U}_{2}} M}+\tfrac{{\sigma}^2}{16 P_{{U}_{1}} M^2}+ \tfrac{{\sigma}^2}{16 P_{{U}_{2}} M^2}
\end{align}
%\emph{Proof}:See Appendix A.
%With the property of positive definite Hermitian matrix
%
%\begin{align}\label{hermitian}
%\text{tr}\{\mathbf{AB}\} \leq \text{tr}\{\mathbf{A}\} \text{tr}\{\mathbf{B}\}
%\end{align}
\subsection{Performance loss analysis}
To satisfy $\text{tr}\{(\mathbf{Q}^{T}\mathbf{Q}^{*})^{-1}\}=1$ and the constant modulus constraint, the training matrix $\mathbf{Q}$ is usually chosen to be a DFT matrix in existing research works in \cite{jensenDFT,zhengCE2}. The DFT matrix is optimal for an IRS with high-resolution or infinite-phase shifters. But for an IRS with low-resolution phase shifters, we propose a Hadamard matrix to replace a DFT matrix. The $N$-points DFT matrix is denoted by $\mathbf{Q}_{DFT}$, which is given by
\begin{align}
[\mathbf{Q}_{DFT}]_{m,n}=e^{-j\tfrac{2\pi(m-1)(n-1)}{M}} ,1\leq m,n\leq N.
\end{align}
What's more, Hadamard matrix is an orthogonal matrix whose entry is given by 1 or -1. An example of $4 \times 4$ Hadamard matrix is as follows
\begin{align}
\mathbf{Q}_{Hadamard}=\left[
  \begin{array}{cccc}
    +1 & +1 & +1 & +1 \\
    +1 & -1 & +1 & -1 \\
    +1 & +1 & -1 & -1 \\
    +1 & -1 & -1 & +1 \\
  \end{array}
\right].
\end{align}
Actually,  a Hadamard matrix only contains 1-bit phase shifter to  achieve an optimal performance. A DFT matrix  requires at least  ${\log}_2 M$-bit phase shifter for each reflection element of the IRS. As $M$ tends to large-scale, this will lead to a high circuit cost.

Assuming IRS adopts phase shifters with $L$ being  the number of discrete phases per phase shifter, each reflection element's phase in $\mathbf{Q}$ takes its nearest value $\tilde{\vartheta}$ from the following set
\begin{align}\label{quanset}
\tilde{\vartheta} \in \bm{\Phi} = \big \{\tfrac{\pi}{L},\tfrac{3\pi}{L},\ldots,\tfrac{(2L-1)\pi}{L} \big \}
\end{align}
which forms the  quantized version $\tilde{\mathbf{Q}}$  of matrix $\mathbf{Q}$. Here, phase quantization noise is assumed to be  uniformly  distributed. Now, let us define the performance loss factor as follows
\begin{align}\label{ql1}
\beta = \tfrac{\text{tr}\{  (\tilde{\mathbf{Q}}^T  \tilde{\mathbf{Q}}^*)^{-1}\}}{\text{tr}\{(\mathbf{Q}^{T}\mathbf{Q}^{*})^{-1}\}}
\end{align}
Note that $\text{tr}\{(\mathbf{Q}^{T}\mathbf{Q}^{*})^{-1}\} =1$. (\ref{ql1}) can be simplified as
\begin{align}\label{ql3}
\beta = \text{tr}\{  (\tilde{\mathbf{Q}}^T  \tilde{\mathbf{Q}}^*)^{-1} \}
\end{align}
When the number of quantization bits is large, $\beta$ can be approximated as
\begin{align}\label{ql2}
\beta \approx  3- 2\text{sinc}(\tfrac{\pi}{L})
\end{align}

\emph{Proof}: See Appendix A.
\section{Simulation results}\label{simulation}

In this section, we perform some numerical simulation to evaluate the performance of IRS-aided TWRN system. System parameters are set as follows: $M= 128$, $K=16$. It is assumed that User1, User2, Relay,  and IRS are in the same plane and the coordinates of User1, User2, Relay, and IRS are (0,0), (100m,0), (0,50m), and (10m,50m), respectively. The path loss exponents of the $\text{User1/User2}\rightarrow \text{R}$, $\text{User1/User2}\rightarrow \text{IRS}$ and $\text{IRS}\rightarrow \text{R}$ are set as 3.5, 2.4, and 2.2, respectively. The path loss at the reference distance of 1~m is chosen as 30~dB. The variance of the noise $\sigma^2$ is -80~dBm. The transmitting powers of User1 and User2 are the same, i.e. $P_{{U}_{1}}=P_{{U}_2}$. The definition of SNR is $P_{{U}_1}/\sigma^2$.

\begin{figure}[htb]
\centering
\includegraphics[width=0.338\textwidth]{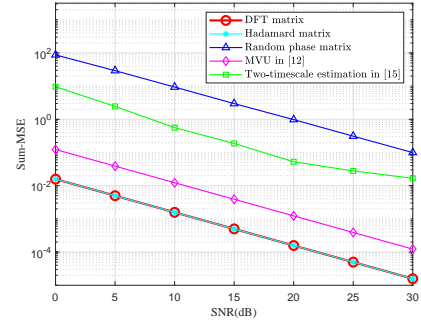}
\centering
\caption{Sum-MSE versus different SNRs with $M$=128.}
\label{fig3}
\end{figure}

Fig.~\ref{fig3} shows the MSE versus SNR with $M=128$ with IRS being random phase matrix (RPM) as a performance benchmark. It is seen that the Hadamard matrix and DFT matrix performs much better than RPM, and achieve the same performance for all values of SNRs because of $\text{tr}\{\mathbf{Q}^* \mathbf{Q}^T\}=M\mathbf{I}_M$. Compared to other channel estimation methods, the sum-MSE of the proposed scheme is smaller than those in \cite{jensenDFT,HuTTCE} due to a higher pilot training overhead in our scheme.

%\begin{figure}[ht]
%\centering
%\includegraphics[width=2.56in]{coverage}
%\centering
%\caption{Objective function value versus number of iterations}
%\end{figure}

%\begin{figure*}[ht]
% \setlength{\abovecaptionskip}{-5pt}
% \setlength{\belowcaptionskip}{-10pt}
% \centering
% \begin{minipage}[t]{0.33\linewidth}
%  \centering
%  \includegraphics[width=2.56in]{PNR.eps}
%  \caption{MSE versus SNR with M=128}
% \end{minipage}%
% \begin{minipage}[t]{0.33\linewidth}
%  \centering
%  \includegraphics[width=2.56in]{quana.eps}
%  \caption{MSE versus quantization bits with different SNRs, M=128}
% \end{minipage}
% \begin{minipage}[t]{0.33\linewidth}
%  \centering
%  \includegraphics[width=2.56in]{quanb.eps}
%  \caption{Quantization loss versus bits with different $M$}
% \end{minipage}
%\end{figure*}

\begin{figure}[htb]
\centering
\includegraphics[width=0.338\textwidth]{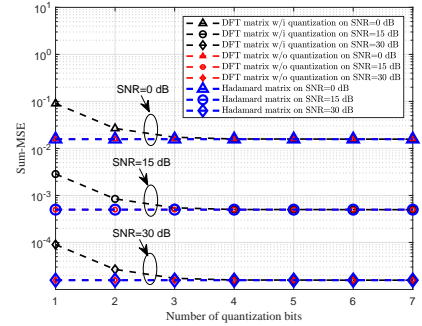}
\centering
\caption{Sum-MSE versus number of quantization  bit with different SNRs, and $M$=128.}
\label{fig4}
\end{figure}
\begin{figure}[htb]
\centering
\includegraphics[width=0.338\textwidth]{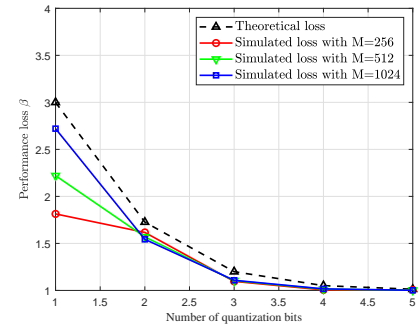}
\centering
\caption{Theoretical and simulated performance loss versus number of quantization  bits with different $M$.}
\label{fig5}
\end{figure}
Fig.~\ref{fig4} shows the Sum-MSE versus the number of phase quantization bits for three typical SNRs 0~dB, 15~dB, and 30~dB and $M=128$. It can be seen that as the number of  quantization bits increases,  the corresponding Sum-MSE approaches the Sum-MSE of infinite precision phase shifter.   the performance loss trend is independent of the values of SNR. Particularly, the channel estimation performance with 3$\sim$4 bit phase shifters is very close to that of the infinite-bit case. This means 3$\sim$4 bits are sufficient for TWRN aided by IRS with finite-bit phase shifters to achieve an omitted performance loss. It is clear that the Hadamard matrix achieves the same Sum-MSE performance as the DFT matrix with infinite bits as the number of quantization bits varies from 1 to 7.

Fig.~\ref{fig5} illustrates the derived theoretical  performance loss factor in (\ref{quan8}) versus the number of  phase quantization bits  with numerical simulated losses for different $M$ as performance references.
%Clearly,  Fig.~\ref{fig5}  has the same trend as Fig.~\ref{fig4}.
The trend in Fig.~\ref{fig5} is consistent with that in Fig.~\ref{fig4}.  As the number of elements of IRS goes to large-scale, the simulated loss factor become closer to the  theoretical  expression derived in (\ref{quan8}). Thus,  this theoretical  expression can be  approximately used to make an analysis of performance loss due to the effect of the number of  phase quantization bits.
\section{Conclusion}\label{conclusion}
In this paper , we have investigated  channel estimation, pilot design and performance loss analysis  of an TWRN aided by IRS with finite-bit phase shifters. To estimate both the direct and  cascaded channels, an excellent pilot pattern was designed,  and the LS channel estimator was presented. Additionally,  the MSE performance loss factor was defined, derived and analyzed.
% From theoretical analysis and simulation results,
%The simulation results showed that
From theoretical analysis and simulation results, if the phase matrix of IRS is chosen to be the Hadamard matrix instead of conventional DFT matrix, then it can achieve an optimal MSE performance in the case of 1-bit phase shifters of IRS. However, the DFT matrix uses 3$\sim$4-bit phase shifter of IRS  also  to achieve  the optimal performance.
\begin{appendices}
\section{Proof of performance loss}
Proof: The quantization error can be defined as $\Delta \mathbf{Q} = \mathbf{Q} - \tilde{\mathbf{Q}}$.
\begin{align}\label{quan1}
&\text{tr}\{  (\tilde{\mathbf{Q}}^T  \tilde{\mathbf{Q}}^*)^{-1}\}=\text{tr}\{\big ( (\mathbf{Q}-\Delta \mathbf{Q})^T  (\mathbf{Q}-\Delta \mathbf{Q})^* \big )^{-1} \} \nonumber \\
&=\text{tr}\{\big ((\mathbf{Q}^T-\Delta \mathbf{Q}^T)(\mathbf{Q}^*-\Delta \mathbf{Q}^*)\big )^{-1} \} =\text{tr}\{ \nonumber \\
&\big (\mathbf{Q}^T\mathbf{Q}^* - \mathbf{Q}^T\Delta \mathbf{Q}^* - \Delta \mathbf{Q}^T\mathbf{Q}^* + \Delta \mathbf{Q}^T\Delta \mathbf{Q}^* \big )^{-1} \}
\end{align}
Note that $\mathbf{Q}^*\mathbf{Q}^T =M\mathbf{I}_M$. Obliviously, $\text{tr}\{(\mathbf{Q}^{T}\mathbf{Q}^{*})^{-1}\} =1$.
Considering that the quantization error at high quantization accuracy becomes extremely small, the inverse of Gram matrix  $\tilde{\mathbf{Q}}^T  \tilde{\mathbf{Q}^*}$ has the following linear approximation
\begin{align}\label{quan2}
 &(\tilde{\mathbf{Q}}^T  \tilde{\mathbf{Q}}^*)^{-1}\approx \big ( M\mathbf{I}_M -\underbrace{ (\mathbf{Q}^T\Delta \mathbf{Q}^* + \Delta \mathbf{Q}^T\mathbf{Q}^*)}_{\Delta \mathbf{I}}  \big )^{-1} \nonumber \\
&=(M(\mathbf{I}_M-\tfrac{1}{M} \Delta \mathbf{I}))^{-1}=\tfrac{1}{M}(\mathbf{I}_M + \tfrac{1}{M} \Delta \mathbf{I} )
\end{align}
which yields
\begin{align}\label{quan3}
\text{tr}\{ (\tilde{\mathbf{Q}}^T \tilde{\mathbf{Q}}^*)^{-1} \} = 1 + \tfrac{1}{M^2} \text{tr} \{ \Delta \mathbf{I} \}
\end{align}
Now, we simplify the second term of the right side of the above equation as follows
\begin{align}\label{quan4}
&\tfrac{1}{M^2}\text{tr}\{\Delta \mathbf{I}\} =\tfrac{2}{M^2}\text{tr}\{\mathbf{Q}^T\Delta \mathbf{Q}^* + \Delta \mathbf{Q}^T\mathbf{Q}^*\}=2\text{tr}\{\mathbf{Q}^T\Delta \mathbf{Q}^*\} \nonumber \\
&=\tfrac{2}{M^2}\text{tr}\{\mathbf{Q}^T(\mathbf{Q}-\mathbf{Q}\odot \Delta \mathbf{Q})^*\} \nonumber \\
&=\tfrac{1}{M^2}\sum_{i=1}^{M}\sum_{m=1}^{M} \mathbf{Q}^*_{(i,m)}(\mathbf{Q}-\mathbf{Q}\odot \Delta \mathbf{Q})^T_{(i,m)} =\tfrac{2}{M^2}\sum_{i=1}^{M}\sum_{m=1}^{M} \nonumber \\
&e^{j\vartheta_{im}}e^{-j\vartheta_{im}}(1-e^{-j\Delta \vartheta_{im}}) =\tfrac{2}{M^2}\sum_{i=1}^{M}\sum_{m=1}^{M}(1-e^{-j\Delta \vartheta_{im}})
\end{align}
Using the above equation and considering that the phase error is assumed to be uniform distribution over the interval [$-\frac{\pi}{L},\frac{\pi}{L}$], we have
\begin{align}\label{quan5}
\tfrac{1}{M^2}\text{tr}(\Delta \mathbf{I})&=2\int^{\tfrac{\pi}{L}}_{-\tfrac{\pi}{L}}p(\Delta \vartheta_{im})(1-e^{-j\Delta \vartheta_{im}})d(\Delta\vartheta_{im})
\end{align}
where
\begin{align}\label{quan6}
p(\Delta \vartheta_{im}) = \tfrac{L}{2\pi}.
\end{align}
Finally, we have
\begin{align}\label{quan7}
2\int^{\tfrac{\pi}{L}}_{-\tfrac{\pi}{L}}\tfrac{L}{2\pi}(1-e^{-j\Delta \vartheta_{im}})d(\vartheta_{im})=2- 2\text{sinc}(\frac{\pi}{L}),
\end{align}
Substituting (\ref{quan7}) in (\ref{quan5}) and then (\ref{quan5}) in (\ref{quan3}) directly gives
\begin{align}\label{quan8}
\text{tr}\{ (\tilde{\mathbf{Q}}^T \tilde{\mathbf{Q}}^*)^{-1} \} \approx 3- 2\text{sinc}(\tfrac{\pi}{L}).
\end{align}
This completes the proof of performance loss factor.
%
%\noindent \emph{Proposition 1}:I don't think you have to give a gift when you first meet someone, so I didn't actually expect to receive any gifts.
%
%\noindent \emph{Proposition 2}:The necklace is okay, but the flowers are not suitable. If a bouquet of flowers I can accept, but in a heart-shaped box is too much like a marriage proposal. The box is also a bit ugly.
%
%\noindent \emph{Proposition 3}:The itinerary is a bit tight. It's more like traveling than dating.
%
%\noindent \emph{Proposition 4}:If you go home late at night, there are also trivial things like removing makeup in the end will not make the girl angry depends mainly on what the girl is feeling, if happy then it is okay.
%
%\noindent \emph{Proposition 5}:If you want to stay in Fujian and she wants to go to Beijing after graduation. This may also be a point of conflict in the future. So she doesn't think it's a good fit in that regard either.
\end{appendices}
\ifCLASSOPTIONcaptionsoff
  \newpage
\fi

\bibliographystyle{IEEEtran}
\bibliography{cit}
\end{document}